\def \SAIT #1 #2 {{\em Mem.\ Soc.\ Astron.\ It.\/} {\bf #1}, #2}
\def \MESS #1 #2 {{\em The Messenger\/} {\bf #1}, #2}
\def \ASTRNACH #1 #2 {{\em Astron. Nach.\/} {\bf #1}, #2}
\def \AAP #1 #2 {{\em Astron. Astrophys.\/} {\bf #1}, #2}
\def \AAL #1 #2 {{\em Astron. Astrophys. Lett.\/} {\bf #1}, L#2}
\def \AAR #1 #2 {{\em Astron. Astrophys. Rev.\/} {\bf #1}, #2}
\def \AAS #1 #2 {{\em Astron. Astrophys. Suppl. Ser.\/} {\bf #1}, #2}
\def \AJ #1 #2 {{\em Astron. J.\/} {\bf #1}, #2}
\def \ANNREV #1 #2 {{\em Ann. Rev. Astron. Astrophys.\/} {\bf #1}, #2}
\def \APJ #1 #2 {{\em Astrophys. J.\/} {\bf #1}, #2}
\def \APJL #1 #2 {{\em Astrophys. J. Lett.\/} {\bf #1}, L#2}
\def \APJS #1 #2 {{\em Astrophys. J. Suppl.\/} {\bf #1}, #2}
\def \APSS #1 #2 {{\em Astrophys. Space Sci.\/} {\bf #1}, #2}
\def \ASR #1 #2 {{\em Adv. Space Res.\/} {\bf #1}, #2}
\def \BAIC #1 #2 {{\em Bull. Astron. Inst. Czechosl.\/} {\bf #1}, #2}
\def \JSQRT #1 #2 {{\em J. Quant. Spectrosc. Radiat. Transfer\/} {\bf #1}, #2}
\def \MN #1 #2 {{\em Mon. Not. R. Astr. Soc.\/} {\bf #1}, #2}
\def \MEM #1 #2 {{\em Mem. R. Astr. Soc.\/} {\bf #1}, #2}
\def \PLR #1 #2 {{\em Phys. Lett. Rev.\/} {\bf #1}, #2}
\def \PASJ #1 #2 {{\em Publ. Astron. Soc. Japan\/} {\bf #1}, #2}
\def \PASP #1 #2 {{\em Publ. Astr. Soc. Pacific\/} {\bf #1}, #2}
\def \NAT #1 #2 {{\em Nature\/} {\bf #1}, #2}

\documentstyle{memsait}
\input epsf.sty
\begin{opening}
\title{CHROMOSPHERIC EMISSION FROM RED GIANTS IN THE OPEN CLUSTER NGC 6940} 
\author{David Barrado y Navascu\'es$^1$ and  A. K. Dupree$^1$}
\institute{$^1$ Harvard--Smithsonian Center for Astrophysics,
Cambridge, MA USA.}
\date{} 
\end{opening}

\begin{document}

\oddpagefooter{}{}{} 
\evenpagefooter{}{}{} 
\ 
\bigskip

\begin{abstract}

The observation of the Ca II H and K lines 
in red giants in  NGC 6940 allows the strength
of chromospheres and their behavior to be evaluated in a population
whose evolution is well understood.
Spectra in the Ca II lines have been obtained for
 giant stars in this cluster.
Emission reversals are present in some objects.
The absolute flux in the chromospheric
emission is determined as a function of effective temperature
for stars on the red giant branch.  The stellar surface flux 
in the Ca II lines decreases smoothly with 
increasing $(B-V)$ in contradiction to model predictions.
\end{abstract}

\section{Introduction}
The decay of magnetic activity in cool stars and the influence of
winds in angular momentum loss 
can be assessed through the strength and profiles of the Ca II lines
($\lambda$3967, $\lambda$3933).
Observation of  Ca II in red giants in cluster stars allows the behavior
of chromospheres to be evaluated in a population
whose evolution is well understood.  Such studies can be used
to test the conjecture that activity is dependent on stellar
effective temperature and mass (e.g. Pasquini \&\ Brocato 1992).
NGC 6940 is an intermediate age cluster containing clump 
giants, bright red giants, and composite binaries.  Age estimates
range from 0.6 Gyr (Carraro \& Chiosi 1994) 
to 1.9 Gyr (Thogerson et al. 1993).

\section{Observations and Calibration}
Echelle spectra in the Ca II lines have been obtained for
74 stars in NGC 6940 using HYDRA (a fiber positioner
with bench spectrograph) at the 4-m telescope at the U.S.
Kitt Peak National Observatory in June 1994. The multi-object
capabilities of this instrument allowed 18 red giants, including
five clump giants,
to be observed simultaneously.  A number of main sequence stars and 
non-member hotter stars were also included.
The targeted red giants generally have a membership probability
$\ge$ 80\% based on proper motions (Sanders 1972) and confirmed by CORAVEL
radial velocities (Mermilliod \&\ Mayor 1989).
The visual magnitude of the stars in this target field 
varied from 
V = 9.2 to V = 13.2, with observed B--V colors ranging
from +0.2 to +1.8.  Fig. 1 shows 
the color-magnitude
diagram of our target stars individually corrected for reddening
(Larsson-Leander 1960), and assuming 
$(m-M)_0 = 10.356$ (Janes \&\ Phelps 1994).  


\begin{figure}
\vspace{8cm}
\includegraphics{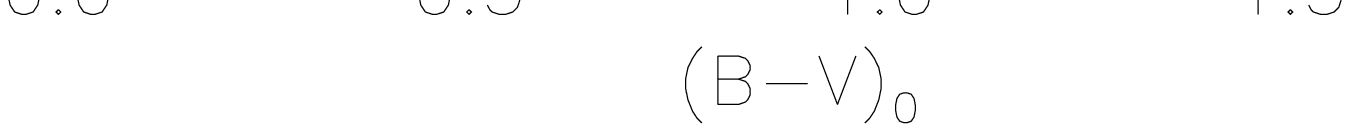}
\caption[h]{Color-magnitude diagram of NGC 6940 indicating the
stars observed by HYDRA. Isochrones for M67 (Tripicco et al. 1993) and
the Hyades [600 Myr (solid curve) and 800 Myr (dash-dot curve);
VandenBerg 1985] are shown.}
\end{figure}

To isolate the Ca II lines,
an order-separating interference filter was used yielding 
a free spectral 
range of 3860--4025\AA; the spectral resolution of 0.17\AA\  
was measured by the FWHM of the Th-Ar comparison
lines.  Spectra were extracted from the CCD images
using the HYDRA packages of IRAF.  Four exposures of 3600 s
each taken on 26 June 1994 
were coadded to produce the final spectra.

Emission cores of the Ca II lines are readily apparent, as is
their dispersion in strength.  For example, two clump stars 
(identified by Mermilliod and Mayor 1989)
VR101\footnote{Here the catalogue number VR is the identification
by Vasilevskis \& Rach (1957).} [$V_0=10.82,(B-V)_0=0.95$] 
and VR108 [$V_0=10.59,(B-V)_0=0.84$],
practically identical in color, have 
emission fluxes differing by  a factor of four (Fig. 2).

The spectra were calibrated by using the absolute
spectrophotometry of Gunn and Stryker (1983).
The ratio
of the H and K emission cores (measured over 2\AA) to a 20\AA\ 
continuum band in our target stars was then normalized to the
Gunn/Stryker fluxes for giant stars of the same value of $(B-V)_0$. 
The emission flux at the star (Fig. 3) was obtained using the
relation between $(B-V)_0$ color and apparent
angular diameter from Barnes et al. (1978).   An extended discussion
of this method is contained in Dupree et al. (1997).

\begin{figure}
\vspace{8cm}
\includegraphics{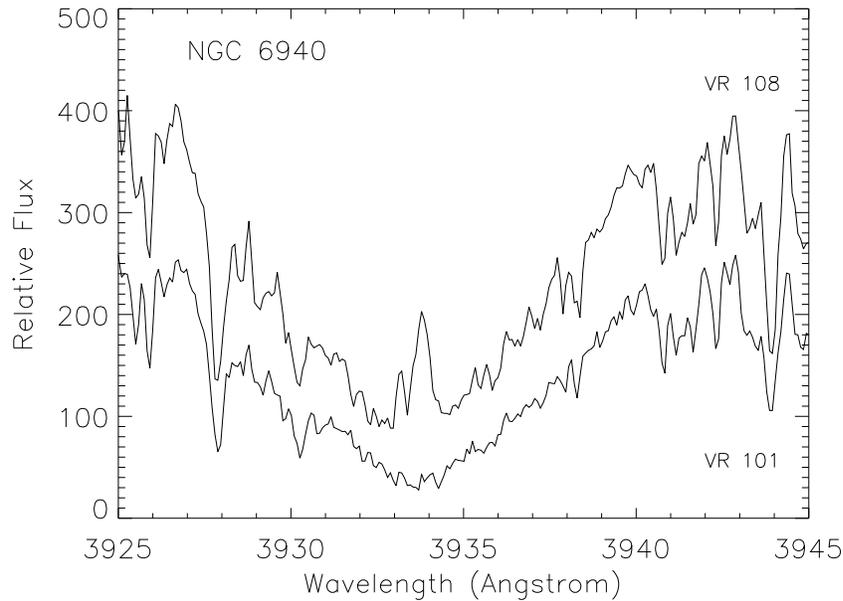}
\caption[h]{Detail of the the Ca K profiles in VR 101 and VR 108.
These clump stars are almost twins, although the chromospheric flux
differs by a factor of 4.}
\end{figure}

\section{Conclusions}

1.  Chromospheres are present in all of the observed red giants,
as indicated by their Ca II emission.

2. The well-developed clump giants possess strong Ca II
emission fluxes indicating 
that chromospheric activity continues into the helium
burning phase.

3. Position in the color magnitude diagram is not a unique
determinant of chromospheric activity as demonstrated by
comparison of the clump stars VR101 and VR108 (Figure 2) which
show a disparity in Ca II flux levels similar to the Hyades
giants.  The spread in  emission level of the Hyades giants
in the chromosphere, transition region and the corona (Stern et al.
1981) has been attributed (Baliunas et al. 1983) 
to the presence of magnetic activity cycles.  These data will
be combined with others in our cluster program to define the
flux levels with stellar evolutionary mass.

4. The smooth gradual decay of Ca II surface flux with decreasing effective
temperature is not in agreement with model predictions.  These
models assume that  angular momentum is conserved, rigid body
rotation, and the moment of inertia changes due to stellar
evolution.  Further studies to define the behavior with
stellar mass, and to detect evidence of mass loss should
place tighter constraints on the models.

\acknowledgements
DBN thanks the Real Colegio
Complutense at Harvard University and the MEC/Fulbright program
for their fellowships.  AKD was a Visiting Astronomer at the 
KPNO to acquire these spectra.

\begin{figure}
\vspace{8cm}
\includegraphics{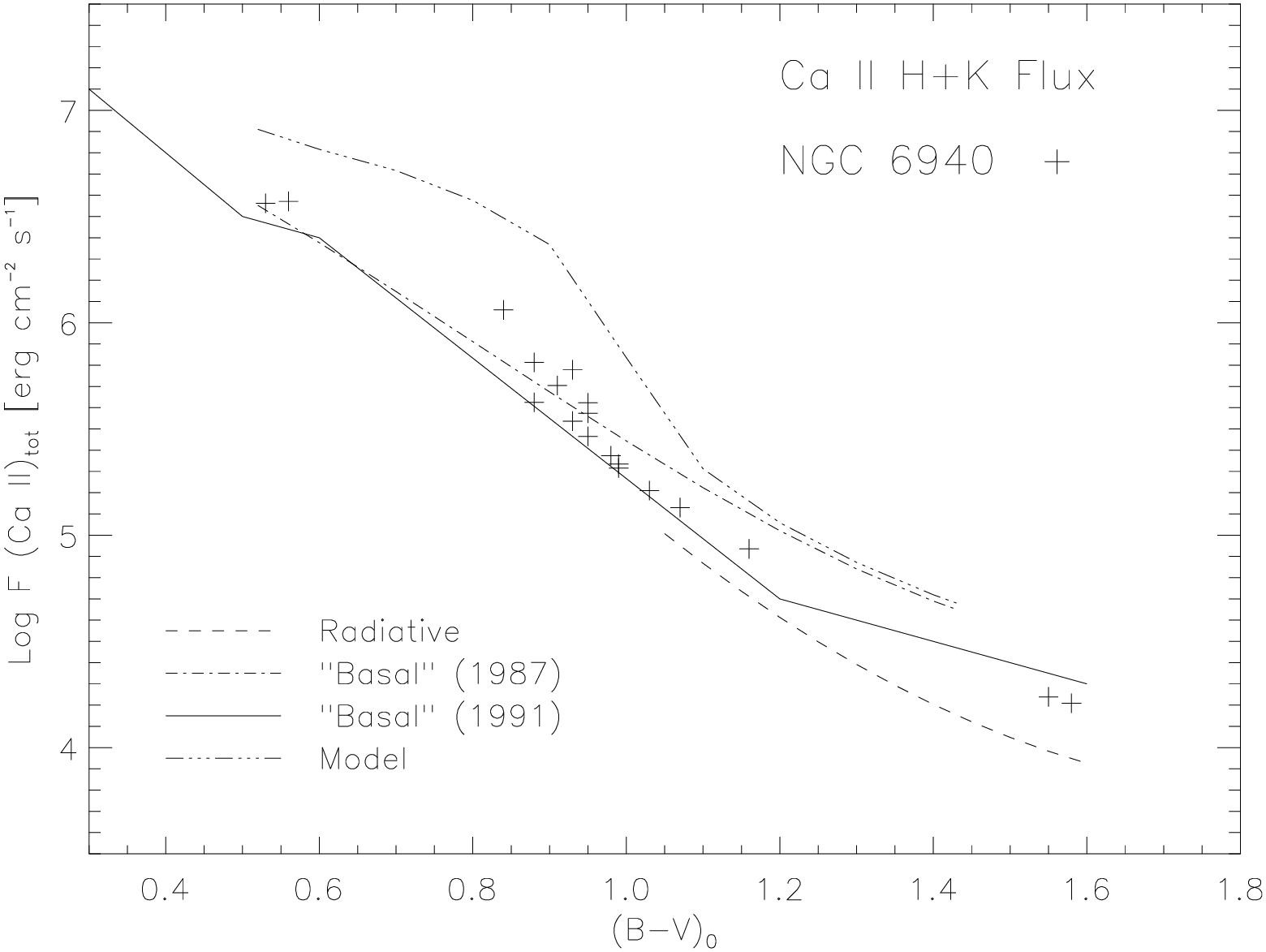}
\caption[h]{The total Ca II ($H+K$) stellar surface flux for the NGC
6940 giants.  the curve marked ``Radiative'' derives from Kurucz
radiative models of giant stars (no chromosphere); the curves
marked ``Basal'' refer to estimates of a ``basal'' flux contribution
(Rutten 1987; Rutten et al. 1991). ``Model'' refers to the predicted
Ca II total flux for a 2$M_\odot$ star (Rutten \&\ Pylyser 1988).}

\end{figure}




\end{document}